\documentclass[twocolumn]{autart}  
\usepackage{amsmath,amssymb,latexsym,amssymb,citesort}

\def \d{{\mathrm{d}}}

\def \pd{\partial}

\def \tl#1{\overset{\kern 1pt\circ}{#1}}
\def \TL#1{\overset{\kern -3pt \circ}{#1}}
\def \TLL#1{\overset{\kern -7pt \circ}{#1}}

\begin{document}

\begin{frontmatter}

\title{Dislocations in the Field Theory of Elastoplasticity} 

\author[ML]{Markus Lazar}
\ead{lazar@mis.mpg.de}
\thanks[]{{\it Present address:} 
	Laboratoire de Mod{\'e}lisation en M{\'e}ca\-nique,
        Universit{\'e} Pierre et Marie Curie, Tour 66,
	4 Place Jussieu, Case 162, F-75252 Paris C{\'e}dex 05, France.}    
\address[ML]{Max-Planck-Institute for Mathematics in the Sciences,\\
             Inselstr. 22-26, D-04103 Leipzig, Germany}                                            

\begin{keyword}                           
dislocations; elastoplasticity; moment stress; stress functions.               
\end{keyword}                             

\begin{abstract}                          
By means of linear theory of elastoplasticity, solutions are given for 
screw and edge dislocations situated in an isotropic solid. 
The force stresses, strain fields, displacements, distortions, dislocation densities
and moment stresses are calculated. The force stresses, strain fields, displacements
and distortions are devoid of singularities predicted by the classical elasticity.
Using the so-called stress function method we found modified stress functions
of screw and edge dislocations.
\end{abstract}

\end{frontmatter}

\section{Introduction}
The traditional description of elastic fields produced by dislocations 
is based on the classical theory of linear elasticity.
However the classical dislocation theory breaks down in the dislocation core region.
The elastic fields due to dislocations which are calculated within the theory of elasticity
contain singularities at the centre of the dislocation. 
This is unfortunate since the dislocation core is an important region in 
metallic plasticity and dislocations and failure occur because of the 
high shear stress action in this region.
Clearly, such singularities
are unphysical and an improved model should eliminate them. 
On the other hand, in conventional plasticity theory 
no internal length scale enters the constitutive relations 
and no size effects are predicted.

A field theory of elastoplasticity is proposed as a means of describing plastic and 
elastic deformation even in the dislocation core.
In this theory the elementary acts of plastic deformation at the microscopical level 
are investigated. In fact, the dislocations are considered as elementary carriers of 
plasticity.
From the field theoretical point of view, dislocations bring new special
degrees of freedom (e.g. anholonomity) 
and their presence leads to a specific response with 
the dimension of a moment stress.
In straightforward manner, a new internal characteristic length scale 
enters the constitutive relation between the dislocation density tensor and the
moment stress tensor.

In this dislocation theory
it turns out that the force stresses, elastic and plastic strains, elastic and plastic distortions
and displacement fields due to dislocations contain no singularities 
and they are finite.
In fact, they vanish at the centre of dislocations. Moreover, the extremum
stress and strain occur at a short distance away from the dislocation line.

The plan of the paper is as follows.
In Section~\ref{elastoplasticity}, I present the basics of the elastoplastic field theory of 
dislocations. 
The non-singular solutions of straight screw and edge dislocations 
calculated in the proposed theory of elastoplasticity are considered
in Section~\ref{screw} and \ref{edge}, respectively. Section~\ref{concl}
concludes the paper.

\section{Theory of elastoplasticity}
\label{elastoplasticity}
In this Section we discuss how the dislocations can be described 
in the frame of field theory of elastoplasticity. We shall in particular 
consider such plastic deformations which originate from the presence of 
dislocations. 

In elastoplasticity the elastic incompatible distortion is given as~\cite{Lazar00,Lazar02a,Lazar02b,Lazar02c}
\begin{align}
\label{dist2}
\beta_{ij}=\pd_j u_i+\phi_{ij}.
\end{align}
It is an additive decomposition of the elastic distortion into compatible and 
purely incompatible distortion. 
This decomposition can be justified by the help of the translational gauge 
theory of defects~\cite{Lazar00,EL88}.
The displacement field $u_i$ gives rise to a compatible distortion 
and the tensor $\phi_{ij}$ is the proper incompatible part of the distortion.
Thus, it is not a formal additive decomposition.
Note that the distortion is dimensionless.
In elastoplasticity the linear elastic strain tensor 
is given by means of the incompatible distortion tensor as
\begin{align}
E_{ij}\equiv\beta_{(ij)}=\frac{1}{2}\big(\pd_i u_j+\pd_ju_i+\phi_{ij}+\phi_{ji}\big),
\quad E_{ij}=E_{ji}.
\end{align}

The (symmetric) force stress is the response quantity (excitation)
with respect to the strain and is given by the generalized Hooke's law
for an isotropic medium
\begin{align}
\sigma_{ij}=
2\mu\left( E_{ij}+\frac{\nu}{1-2\nu}\,\delta_{ij} E_{kk}\right),\quad
\sigma_{ij}=\sigma_{ji},
\end{align}
where $\mu$, $\nu$ are shear modulus and Poisson's ration, respectively.
The (symmetric) force stress has to satisfy the force equilibrium condition
\begin{align}
\label{FEq}
\pd_i\sigma_{ij}=0=\pd_j\sigma_{ij}.
\end{align}
The skew-symmetric part of the distortion tensor defines
the elastic rotation of a dislocation~\cite{deWit73a,deWit73b}
\begin{align}
\omega_i\equiv-\frac{1}{2}\,\epsilon_{ijk}\beta_{jk}.
\end{align}
The rotation vector gives rise to a rotation gradient (deWit's bend-twist tensor)
\begin{align}
k_{ij}=\pd_j\omega_i.
\end{align}

The total strain $E^T_{ij}$ and distortion $\beta^T_{ij}$ are defined in 
terms of the displacement field $u_i$. They must be compatible. 
In the presence of dislocations, the total strain is not completely elastic,
but a part of it is plastic~\cite{deWit73a} so that
\begin{align}
&E^T_{ij}\equiv\pd_{(i}u_{j)}=E_{ij}+E^P_{ij},\\
&\beta^T_{ij}\equiv\pd_j u_i=\beta_{ij}+\beta^P_{ij},\quad
\beta^P_{ij}\equiv-\phi_{ij}.\nonumber
\label{total-strain}
\end{align}
Here $E^P_{ij}=-\phi_{(ij)}$ is the plastic strain and $\beta^P_{ij}$ the plastic distortion 
which are not derivable from the displacement field when the plastic strain
is incompatible. 
In performing the differentiations of the displacement~$u_i$ 
we obtain the elastic distortion $\beta_{ij}$ plus
excess terms which we identify with the plastic distortion $\beta^P_{ij}$.
We notice that this decomposition of $E_{ij}^T$ is close to the Green-Naghdi 
decomposition (see~\cite{GN}).

The dislocation density tensor is defined by means of the distortion
tensor 
\begin{align}
T_{ijk}:=\pd_j\beta_{ik}-\pd_k\beta_{ij}=
\pd_j\phi_{ik}-\pd_k\phi_{ij},\quad T_{ijk}=-T_{ikj},
\end{align}
and
\begin{align}
\label{torsion2}
T_{ijk}=-\pd_j\beta^P_{ik}+\pd_k\beta^P_{ij}.
\end{align}
Physically, the incompatible distortion corresponds to dislocations.
The dislocation density tensor has the dimension of an inverse length.
It is a fundamental quantity in plasticity because the
dislocation is the elementary carrier of plasticity.
The usual dislocation density tensor $\alpha_{ij}$  is recovered by
(see also~\cite{Kroener81})
\begin{align}
\label{alpha-ij}
\alpha_{ij}:=\frac{1}{2}\,\epsilon_{jkl}T_{ikl}
            =\epsilon_{jkl}\pd_k \beta_{il} =-\epsilon_{jkl}\pd_k \beta^P_{il}.
\end{align}
Here the index $i$ indicates the direction of the Burgers vector, 
$j$ the dislocation line direction. 
Thus, the diagonal components of $\alpha_{ij}$ represent screw dislocations,
the off-diagonal components edge dislocations.
In this approach the dislocation density is not given a priori as a delta function.
It follows from the physics of the problem.
Eqs.~(\ref{torsion2}) and (\ref{alpha-ij}) justify the identification of 
$-\phi_{ij}$ with the plastic distortion.
The dislocation density tensor satisfies the translational Bianchi identity
\begin{align}
\label{bianchi}
\pd_j \alpha_{ij}=0,
\end{align}
which means that dislocations do not end inside the medium.

The moment stress which is the response quantity (excitation) to the dislocation density 
is given by (see~\cite{Lazar02c})
\begin{align}
\label{moment2}
&H_{ijk}=\frac{a_1}{2}\Big( T_{ijk}-T_{jki}-T_{kij}\\
&\hspace{2.5cm}+\frac{2\nu}{1-\nu}\,\big(\delta_{ij}T_{llk}+\delta_{ik}T_{ljl}
\big)\Big),\nonumber\\ 
&H_{ijk}=-H_{ikj}\nonumber
\end{align}
or with $H_{ij}=\frac{1}{2}\epsilon_{jkl} H_{ikl}$
\begin{align}
\label{moment3}
H_{ij}=a_1\left\{\frac{1}{1-\nu}\, \alpha_{ij} -\frac{\nu}{1-\nu}\,\alpha_{ji}
-\frac{1}{2}\,\delta_{ij}\alpha_{kk}\right\}.
\end{align}
Obviously, (\ref{moment2}) and (\ref{moment3}) are (linear) constitutive relations
between dislocation density and moment stress in an isotropic medium.
The coefficient $a_1$ has the dimension of a force.

The field equation in elastoplasticity for the force stress in an isotropic medium 
is proposed as 
the following inhomogeneous Helmholtz equation~\cite{Lazar02c}
\begin{align}
\label{stress-fe}
\big(1-\kappa^{-2}\Delta\big)\sigma_{ij}=\tl\sigma {}_{ij},\qquad \kappa^2=\frac{2\mu}{a_1},
\end{align}
where $\tl\sigma {}_{ij}$ is the stress tensor
obtained for the same traction boundary-value problem 
within the ``classical'' theory of dislocations.
In the field theory of elastoplasticity
Eq.~(\ref{stress-fe}) is obtained from the moment stress equilibrium 
condition (see~\cite{Lazar02c}).
It is important to note that (\ref{stress-fe}) agrees with the field equation
for the stress field in Eringen's nonlocal elasticity~\cite{Eringen83,Eringen85}
and in gradient elasticity~\cite{GA99}.
The factor $\kappa^{-1}$ has the physical dimension of a length and 
therefore it defines an internal characteristic length 
(dislocation length scale).
Using the inverse of the generalized Hooke's law and (\ref{stress-fe}) we obtain an inhomogeneous
Helmholtz equation for the strain fields (see~\cite{Lazar02c})
\begin{align}
\label{strain-fe}
\big(1-\kappa^{-2}\Delta\big)E_{ij}=\tl E {}_{ij},
\end{align}
where $\tl E {}_{ij}$ is the classical strain tensor.
Equation~(\ref{strain-fe}) is similar to the equation for strain of the 
gradient theory used by Gutkin and Aifantis~\cite{GA96,GA97,GA99}
if we identify $\kappa^{-2}$ with their corresponding gradient coefficient
(see, e.g., equation~(4) in~\cite{GA99}).
From the fact that the field equation of the force stress~(\ref{stress-fe}) 
has the same form in elastoplastic field theory, strain gradient elasticity and nonlocal elasticity
some relations for the corresponding solutions will follow.
We assume that the stress and strain fields at infinity should have the same form
for both the classical and elastoplastic field theory.

\section{Screw dislocation in elastoplasticity}
\label{screw}
Consider an infinitely long screw dislocation whose dislocation line 
coincides with the $z$-axis of a Cartesian coordinate system.
Due to the symmetry of the problem, we choose the Burgers vector, 
which is parallel to the dislocation line, 
in $z$-direction: $b_x=b_y=0$, $b_z=b$.
We solve the force stress equilibrium condition~(\ref{FEq}) identically by a
so-called stress function ansatz.
Using the stress function ansatz~\cite{Kroener81,Kroener58,HL} 
the elastic stress of a Volterra screw dislocation, 
in Cartesian coordinates, reads 
\begin{align}
&\tl\sigma_{xz}=
\tl\sigma_{zx}=-\pd_y \Phi=-\frac{\mu b}{2\pi}\,\frac{y}{r^2},
\nonumber\\
&\tl\sigma_{yz}=
\tl\sigma_{zy}=\pd_x\Phi=\frac{\mu b}{2\pi}\,\frac{x}{r^2},
\end{align}
or, in cylindrical coordinates,
\begin{align}
\label{T-bg}
\tl\sigma_{z\varphi}=
\tl\sigma_{\varphi z}=\pd_r\Phi=\frac{\mu b}{2\pi r},
\end{align}
where $r^2=x^2+y^2$ and $\varphi=\arctan y/x$.
Here, $\Phi$ is the well-known stress function of elastic torsion, sometimes 
called Prandtl's stress function. It is given by (see, e.g.,~\cite{Kroener81})
\begin{align}
\Phi=\frac{\mu b}{2\pi}\, \ln r.
\end{align}
This is Green's function of the two-dimensional potential equation
\begin{align}
\Delta \Phi=\mu b\, \delta(r).
\end{align}
Here $\delta(r)$ is the two-dimensional Dirac delta function
and $\Delta$ denotes the two-dimensional Laplacian $\pd^2_{xx}+\pd^2_{yy}$.
Obviously, the ``classical'' stress fields are singular at the dislocation line.
For the modified stress we make the ansatz
\begin{align}
\label{}
\sigma_{ij}=
\left(\begin{array}{ccc}
0 & 0 & -\pd_{y} F \\
0 & 0 & \pd_{x} F\\
-\pd_{y} F  &\pd_{x} F  & 0
\end{array}\right),
\end{align}
where $F$ is called the modified Prandtl stress function.
Substituting the stress functions into (\ref{stress-fe}), 
we get the following inhomogeneous Helmholtz equation
\begin{align}
\left(1-\kappa^{-2}\Delta\right)F=\frac{\mu b}{2\pi}\, \ln r.
\end{align} 
The solution of the modified stress function of a screw dislocation
is given by
\begin{align}
F=\frac{\mu b}{2\pi}\Big\{\ln r +K_0(\kappa r)\Big\},
\end{align}
where $K_n$ is the modified Bessel function of the second kind and of 
order $n$.
Consequently, we find the force stresses
\begin{align}
\label{T-cart}
&\sigma_{xz}
=-\frac{\mu b}{2\pi}\,\frac{y}{r^2}\Big\{1-\kappa r K_1(\kappa r)\Big\},
\nonumber\\
&\sigma_{yz}=
\frac{\mu b}{2\pi}\,\frac{x}{r^2}\Big\{1-\kappa r K_1(\kappa r)\Big\},
\end{align}
and in cylindrical coordinates
\begin{align}
\label{T-cyl}
\sigma_{\varphi z}=\frac{\mu b}{2\pi}\,\frac{1}{r}\Big\{1-\kappa r K_1(\kappa r)\Big\}.
\end{align}
The appearance of the modified Bessel function in (\ref{T-cart}) and (\ref{T-cyl})
leads to the elimination of classical singularity $\sim r^{-1}$ at the dislocation line.
The modified stress field (\ref{T-cart}) agrees
with the stress field calculated by Eringen~\cite{Eringen83,Eringen85}
within his version of nonlocal elasticity. 
Additionally, 
it is interesting to note that the stress field~(\ref{T-cart})
is the same as the one obtained by Gutkin and Aifantis~\cite{GA99}
in their version of  gradient elasticity.
The stress $\sigma_{yz}$ has its extreme value 
$|\sigma_{yz}(x,0)|\simeq 0.399\kappa \frac{\mu b}{2\pi}$ at 
$|x|\simeq 1.114 \kappa^{-1}$,
whereas the stress $\sigma_{xz}$ has its extreme value 
$|\sigma_{xz}(0,y)|\simeq 0.399\kappa \frac{\mu b}{2\pi}$ at 
$|y|\simeq 1.114 \kappa^{-1}$.
We notice that the extremum stress may be identified with the 
theoretical shear strength.
The factor $\kappa$ should be fitted by comparing predictions of the theory 
with experimental results. 
In general, the parameter $\kappa$ can be used to determine the width of a dislocation 
and the amplitude of the force stress.

Let us now calculate the distortion of a screw dislocation.
The distortion $\beta_{ij}$ is given in terms of the stress function
\begin{align}
\label{dist-ansatz-screw}
\beta_{ij}=\!\frac{1}{2\mu}
\!\left(\begin{array}{ccc}
0 & 0 & \!\!-\pd_{y} F +2\mu\omega_1\\
0 & 0 & \!\!\pd_{x} F-2\mu\omega_2\\
-\pd_{y} F -2\mu\omega_1 &\pd_{x} F+2\mu\omega_2  &\!\! 0
\end{array}\!\right),
\end{align}
where the two functions $\omega_1$ and $\omega_2$ are used to express the 
antisymmetric part of the distortion ($\beta_{[xz]}\equiv \omega_1$ and 
$\beta_{[yz]}\equiv -\omega_2$).
Eventually, $\omega_1$ and $\omega_2$  are determined from the conditions
\begin{align}
\alpha_{xy}=T_{xzx}&=-\frac{1}{2\mu}\, \pd_x\big(2\mu\omega_1-\pd_y F\big)\equiv 0,\nonumber\\
\alpha_{yx}=T_{yyz}&=-\frac{1}{2\mu}\, \pd_y\big(2\mu\omega_2-\pd_x F\big)\equiv 0.
\end{align}
One finds for the distortion tensor of the screw dislocation 
\begin{align}
\label{dist-screw}
&\beta_{zx}=-\frac{b}{2\pi}\,\frac{y}{r^2}\Big\{1-\kappa r K_1(\kappa r)\Big\},
\nonumber\\
&\beta_{zy}=\frac{b}{2\pi}\,\frac{x}{r^2}\Big\{1-\kappa r K_1(\kappa r)\Big\},
\end{align}
and for elastic rotation vector
\begin{align}
&\omega_x\equiv\omega_2=\frac{b}{4\pi}\,\frac{x}{r^2}\Big\{1-\kappa r K_1(\kappa r)\Big\},
\nonumber\\
&\omega_y\equiv\omega_1=\frac{b}{4\pi}\,\frac{y}{r^2}\Big\{1-\kappa r K_1(\kappa r)\Big\}.
\end{align}
The rotation vector is in agreement with the result calculated in the 
Cosserat theory~\cite{Nowacki73}.
The incompatible elastic strain reads
\begin{align}
&E_{xz}=-\frac{b}{4\pi}\,\frac{y}{r^2}\Big\{1-\kappa r K_1(\kappa r)\Big\},
\nonumber\\ 
&E_{yz}=\frac{b}{4\pi}\,\frac{x}{r^2}\Big\{1-\kappa r K_1(\kappa r)\Big\}.
\end{align}

By means of the distortion tensor~(\ref{dist-screw}) 
the effective Burgers vector, $b_i(r)=\oint_\gamma\beta_{ij}\d x_j$, 
for a circular circuit of radius $r$ is given by
\begin{align}
\label{Burger-screw}
b_z(r)=\oint_\gamma\big(\beta_{zx}\d x+\beta_{zy}\d y\big)
      =b\Big\{1-\kappa r K_1(\kappa r)\Big\},
\end{align}
where $\gamma$ is the Burgers circuit. 
It depends on the radius $r$. In fact, we find $b_z(0)=0$ and $b_z(\infty)=b$.
This effective Burgers vector differs appreciably from the constant 
value $b$ in the core region from $r=0$ up to $r\simeq 6\kappa^{-1}$. 
Thus, it is suggestive to take $r_c\simeq 6\kappa^{-1}$ as the characteristic length 
(dislocation core radius). 
Therefore, in the field theory of elastoplasticity the dislocation has a core in 
quite natural manner.	
Outside this core region the Burgers vector reaches its constant value. 
Accordingly, the classical and the elastoplastic solution coincide outside 
the core region.

If we use the decomposition~(\ref{dist2}) of the distortion 
into the compatible and purely incompatible distortion,
the displacement field of a screw dislocation turns out to be 
(see also~\cite{Edelen96})
\begin{align}
\label{u-T}
u_z=\frac{b}{2\pi}\big(1-\kappa r K_1(\kappa r)\big)\varphi,
\end{align}
where $\varphi$ is multi-valued.
Thus, the proper incompatible part of the distortion 
is the (negative) plastic distortion which is confined in the 
dislocation core region 
\begin{align}
\label{plastic-dist-screw}
\phi_{zx}
=-\frac{b\kappa^2}{2\pi}\,x \varphi K_0(\kappa r),
\quad
\phi_{zy}
=-\frac{b\kappa^2}{2\pi}\,y \varphi K_0(\kappa r).
\end{align}
It fulfils Eq.~(\ref{alpha-ij}).
The plastic strain reads
\begin{align}
E^P_{zx}=-\frac{1}{2}\, \phi_{zx},
\qquad
E^P_{zy}=-\frac{1}{2}\, \phi_{zy}.
\end{align}

Now we are able to calculate the dislocation density by means of the
distortion tensor. We obtain
\begin{align}
\label{alpha-zz}
\alpha_{zz}=T_{zxy}
=\frac{1}{\mu}\, \Delta F=\frac{b\kappa^2}{2\pi}\, K_0(\kappa r).
\end{align}
Of course, this dislocation density satisfies the condition~(\ref{bianchi}).
In the limit as $\kappa^{-1}\rightarrow 0$, the elastoplastic 
result~(\ref{alpha-zz}) converts to the classical dislocation density
$\alpha_{zz}=b\,\delta(r)$.
By the help of (\ref{moment3})
the localised moment stresses caused by the screw dislocation 
can be expressed in terms of the dislocation density as
\begin{align}
&H_{xx}=-\frac{\mu b}{2\pi}\, K_0(\kappa r),\quad
H_{yy}=-\frac{\mu b}{2\pi}\, K_0(\kappa r),\nonumber\\
&H_{zz}=\frac{\mu b}{2\pi}\, K_0(\kappa r),
\end{align}
and 
\begin{align}
H_{kk}=-\frac{\mu b}{2\pi}\, K_0(\kappa r).
\end{align}
Accordingly, moment stresses of twisting-type occur at all positions where 
the dislocation density $\alpha_{zz}$ is non-vanishing.
When $\kappa^{-1}\rightarrow 0$, the moment stresses vanish. 

The far-reaching rotation gradients of a screw dislocation read
\begin{align}
k_{xx}&\!=\!\frac{b}{4\pi r^4}\Big\{\!\big(y^2-x^2\big)\!\big(1-\kappa r K_1(\kappa r)\big)
                \!+\!\kappa^2 x^2 r^2 K_0(\kappa r)\!\Big\},\nonumber \\
k_{yy}&\!=\!\frac{b}{4\pi r^4}\Big\{\!\big(x^2-y^2\big)\!\big(1-\kappa r K_1(\kappa r)\big)
                \!+\!\kappa^2 y^2 r^2 K_0(\kappa r)\!\Big\},\nonumber\\
k_{xy}&=-\frac{b}{4\pi r^4}\, xy \Big\{2\big(1-\kappa r K_1(\kappa r)\big)
                -\kappa^2 r^2 K_0(\kappa r)\Big\},\nonumber\\
k_{yx}&=k_{xy},\nonumber\\
k_{jj}&=\frac{1}{2}\,\alpha_{zz}.
\end{align}
They are in agreement with the expressions calculated within the theory of Cosserat 
media (see~\cite{Kessel70,Nowacki74,Minagawa77}).

\section{Edge dislocation in elastoplasticity}
\label{edge}
Consider an infinitely long edge dislocation whose dislocation line 
coincides with the $z$-axis while the Burgers vector
is parallel to the $x$-axis: $b_x=b$, $b_y=b_z=0$.
The extra half plane lies in the plane $x=0$. 
In order to satisfy the equilibrium condition~(\ref{FEq})
we use the second order stress function $f$
and specialize to the plane problem of an edge dislocation by
setting $\pd_z\equiv 0$. 
The classical stress field of a straight edge dislocation 
can be given in terms of Airy's stress function according to
\begin{align}
\label{stress-ansatz2}
\tl\sigma {}_{ij}=
\left(\begin{array}{ccc}
\pd^2_{yy}\chi & -\pd^2_{xy}\chi & 0\\
-\pd^2_{xy}\chi & \pd^2_{xx}\chi & 0\\
0& 0& \nu\Delta\chi
\end{array}\right).
\end{align}
Airy's stress function~\cite{Kroener58,Kroener81}
\begin{align}
\label{Airy1}
\chi=-A\, y\ln r,\qquad A=\frac{\mu b}{2\pi(1-\nu)},
\end{align}
fulfils the following inhomogeneous bipotential (or biharmonic) equation 
\begin{align}
\Delta\Delta\,\chi=-4\pi A\,\pd_y\delta(r).
\end{align}
For the modified stress
we make the following stress function ansatz 
\begin{align}
\label{stress-ansatz}
\sigma_{ij}=
\left(\begin{array}{ccc}
\pd^2_{yy}f & -\pd^2_{xy}f & 0\\
-\pd^2_{xy}f & \pd^2_{xx}f & 0\\
0& 0& \nu\Delta f
\end{array}\right).
\end{align}
In addition, the strain is given in terms of the stress function as
\begin{align}
\label{strain-ansatz}
E_{ij}=\frac{1}{2\mu}
\left(\begin{array}{ccc}
\pd^2_{yy}f-\nu\Delta f & -\pd^2_{xy}f & 0\\
-\pd^2_{xy}f & \pd^2_{xx}f-\nu\Delta f & 0\\
0& 0& 0
\end{array}\right).
\end{align}
Substituting (\ref{stress-ansatz}) and (\ref{stress-ansatz2})
into (\ref{stress-fe}) we get the inhomogeneous Helmholtz equation 
\begin{align}
\label{f_fe}
\Big(1-\kappa^{-2}\Delta\Big)f=- A\, y\ln r .
\end{align}
The solution of the modified stress function of a straight edge dislocation
is given by~\cite{Lazar02c}
\begin{align}
\label{f_edge1}
f=-\frac{\mu b}{2\pi(1-\nu)}\, y \bigg\{\ln r 
+\frac{2}{\kappa^2 r^2}\Big(1-\kappa r K_1(\kappa r)\Big)\bigg\}, 
\end{align}
where the first piece is Airy's stress function. 

By means of Eqs.~(\ref{stress-ansatz}) and (\ref{f_edge1}),
the modified stress of a straight edge dislocation is given as
\begin{align}
&\sigma_{xx}=-\frac{\mu b}{2\pi(1-\nu)}\, 
\frac{y}{r^4}\bigg\{\big(y^2+3x^2\big)+\frac{4}{\kappa^2r^2}\big(y^2-3x^2\big)\nonumber\\
&\hspace{2cm}
-2 y^2\kappa r K_1(\kappa r)-2\big(y^2-3x^2\big) K_2(\kappa r)\bigg\},\nonumber\\
&\sigma_{yy}=-\frac{\mu b}{2\pi(1-\nu)}\, 
\frac{y}{r^4}\bigg\{\big(y^2-x^2\big)-\frac{4}{\kappa^2r^2}\big(y^2-3x^2\big)\nonumber\\
&\hspace{2cm}
-2 x^2\kappa r K_1(\kappa r)+2\big(y^2-3x^2\big) K_2(\kappa r)\bigg\},\nonumber\\
&\sigma_{xy}=\frac{\mu b}{2\pi(1-\nu)}\, 
\frac{x}{r^4}\bigg\{\big(x^2-y^2\big)-\frac{4}{\kappa^2r^2}\big(x^2-3y^2\big)\nonumber\\
&\hspace{2cm}
-2 y^2\kappa r K_1(\kappa r)+2\big(x^2-3y^2\big) K_2(\kappa r)\bigg\},\nonumber\\
\label{T_zz}
&\sigma_{zz}=-\frac{\mu b\nu }{\pi(1-\nu)}\, 
\frac{y}{r^2}\Big\{1-\kappa r K_1(\kappa r)\Big\}.
\end{align}
The trace of the stress tensor $\sigma_{kk}$ produced by the edge dislocation in
an isotropic medium is
\begin{align}
\label{hyd_p}
\sigma_{kk}=-\frac{\mu b(1+\nu)}{\pi(1-\nu)}\, 
\frac{y}{r^2}\Big\{1-\kappa  r K_1(\kappa r)\Big\}.
\end{align}

Let us now discuss some details of the stress fields in the core region.
The stress fields have no artificial singularities at the core  
and the extremum stress occurs at a short distance away from 
the dislocation line. 
In fact, when $r\rightarrow 0$, we have
\begin{align}
K_1(\kappa r)\rightarrow \frac{1}{\kappa r},\qquad
K_2(\kappa r)\rightarrow -\frac{1}{2}+\frac{2}{(\kappa r)^2},
\end{align}
and thus $\sigma_{ij}\rightarrow 0$.
It can be seen that the stresses have the
following extreme values:
$|\sigma_{xx}(0,y)|\simeq 0.546\kappa \frac{\mu b}{2\pi(1-\nu)}$ at 
$|y|\simeq 0.996 \kappa^{-1}$,
$|\sigma_{yy}(0,y)|\simeq 0.260 \kappa
\\ 
\frac{\mu b}{2\pi(1-\nu)}$ at 
$|y|\simeq 1.494 \kappa^{-1}$,
$|\sigma_{xy}(x,0)|\simeq 0.260 \kappa \frac{\mu b}{2\pi(1-\nu)}$ at 
$|x|\simeq 1.494 \kappa^{-1}$,
and
$|\sigma_{zz}(0,y)|\simeq 0.399\kappa \frac{\mu b\nu}{\pi(1-\nu)}$ at 
$|y|\simeq 1.114 \kappa^{-1}$.
The stresses $\sigma_{xx}$,  $\sigma_{yy}$ and  $\sigma_{xy}$ are modified 
near the dislocation core ($0\le r\le 12\kappa^{-1}$). 
Note that $|\sigma_{xy}(x,0)|\simeq 0.260 \kappa\frac{\mu b}{2\pi(1-\nu)}$
can be identified with the theoretical shear strength.
The stresses
$\sigma_{zz}$ and $\sigma_{kk}$ are modified
in the region: $0\le r\le 6\kappa^{-1}$.
Far from the dislocation line ($r\gg 12\kappa^{-1}$) the elastoplastic 
and the classical solutions of the stress of an edge dislocation coincide.
Thus, the characteristic internal length $\kappa^{-1}$ determines the
position and the magnitude of the stress extrema.
It is interesting and important to note that the 
solution~(\ref{T_zz}) agrees precisely 
with the gradient solution given by Gutkin and 
Aifantis~\cite{GA99} 
(with $\kappa^{-2}=c$, $c$ is the gradient coefficient).

For the elastic strain of an edge dislocation we find
\begin{align}
&\!E_{xx}\!=\!-\frac{b}{4\pi(1-\nu)} 
\frac{y}{r^2}
\bigg\{\!(1-2\nu)+\frac{2x^2}{r^2}+\frac{4}{\kappa^2r^4}\big(y^2-3x^2\big)
\nonumber\\
&\hspace{0.6cm}-2\left(\frac{y^2}{r^2}-\nu\right)\kappa r K_1(\kappa r)
-\frac{2}{r^2}\big(y^2-3x^2\big) K_2(\kappa r)\bigg\},\nonumber\\
&\!E_{yy}\!=\!-\frac{b}{4\pi(1-\nu)} 
\frac{y}{r^2}
\bigg\{\!(1-2\nu)-\frac{2x^2}{r^2}-\frac{4}{\kappa^2r^4}\big(y^2-3x^2\big)\nonumber\\
&\hspace{0.6cm}-2\left(\frac{x^2}{r^2}-\nu\right)\kappa r K_1(\kappa r)
+\frac{2}{r^2}\big(y^2-3x^2\big) K_2(\kappa r)\bigg\},\nonumber\\
&\!E_{xy}=\frac{b}{4\pi(1-\nu)}\, 
\frac{x}{r^2}
\bigg\{1-\frac{2y^2}{r^2}-\frac{4}{\kappa^2r^4}\big(x^2-3y^2\big)\nonumber\\
&\hspace{0.6cm} -\frac{2y^2}{r^2}\,\kappa r K_1(\kappa r)
+\frac{2}{r^2}\big(x^2-3y^2\big) K_2(\kappa r)\bigg\},
\end{align}
which is in agreement with the solution given by Gutkin and Aifantis~\cite{GA97,GA99}
in the framework of strain gradient elasticity.
The plane-strain condition $E_{zz}=0$ of the classical 
dislocation theory is valid. 
The components of the strain tensor have the
following extreme values ($\nu=0.3$):
$|E_{xx}(0,y)|\simeq 0.308\kappa \frac{ b}{4\pi(1-\nu)}$ at 
$|y|\simeq 0.922 \kappa^{-1}$,
$|E_{yy}(0,y)|\simeq 0.010 \kappa\frac{ b}{4\pi(1-\nu)}$ at 
$|y|\simeq 0.218 \kappa^{-1}$, 
$|E_{yy}(0,y)|\simeq 0.054 \kappa\frac{ b}{4\pi(1-\nu)}$ at 
$|y|\simeq 4.130 \kappa^{-1}$, 
and
$|E_{xy}(x,0)|\simeq 0.260 \kappa\frac{ b}{4\pi(1-\nu)}$ at 
$|x|\simeq 1.494 \kappa^{-1}$. 
It is interesting to note that 
$E_{yy}(0,y)$ is much smaller than $E_{xx}(0,y)$ within
the core region (see also~\cite{GA99}).
The dilatation $E_{kk}$ reads
\begin{align}
E_{kk}=-\frac{b(1-2\nu)}{2\pi(1-\nu)}\, 
\frac{y}{r^2}\Big\{1-\kappa  r K_1(\kappa r)\Big\}.
\end{align}

Now we calculate the distortion of an edge dislocation.
The distortion $\beta_{ij}$ is given in terms of the stress function~(\ref{f_edge1}):
\begin{align}
\label{dist-ansatz-edge}
\beta_{ij}=\frac{1}{2\mu}
\left(\begin{array}{ccc}
\pd^2_{yy}f-\nu\Delta f & -\pd^2_{xy}f +2\mu\omega& 0\\
-\pd^2_{xy}f-2\mu\omega & \pd^2_{xx}f-\nu\Delta f & 0\\
0& 0& 0
\end{array}\right),
\end{align}
where $\omega$ is used to express the antisymmetric part of the distortion,
$\omega\equiv\beta_{[xy]}$.
Eventually, $\omega$ is determined from the conditions:
\begin{align}
\alpha_{xz}&=T_{xxy}=\frac{1}{2\mu}\big(2\mu\pd_x\omega-(1-\nu)\pd_y\Delta f\big),\nonumber\\
\alpha_{yz}&=T_{yxy}=\frac{1}{2\mu}\big(2\mu\pd_y\omega+(1-\nu)\pd_x\Delta f\big)\equiv 0.
\end{align}
We find for the elastic distortion of the edge dislocation
\begin{align}
&\!\beta_{xx}\!=\!-\frac{b}{4\pi(1-\nu)} 
\frac{y}{r^2}
\bigg\{\!(1-2\nu)+\frac{2x^2}{r^2}+\frac{4}{\kappa^2r^4}\big(y^2-3x^2\big)\nonumber\\
&\hspace{0.5cm}
-2\left(\frac{y^2}{r^2}-\nu\right)\kappa r K_1(\kappa r)
-\frac{2}{r^2}\big(y^2-3x^2\big) K_2(\kappa r)\bigg\},\nonumber\\
&\!\beta_{xy}\!=\!\frac{b}{4\pi(1-\nu)} \,
\frac{x}{r^2}
\bigg\{\!(3-2\nu)-\frac{2y^2}{r^2}-\frac{4}{\kappa^2r^4}\big(x^2-3y^2\big)\nonumber\\
&\hspace{0.0cm} 
-2\!\left(\!(1-\nu)+\frac{y^2}{r^2}\right)\!\kappa r K_1(\kappa r)
+\frac{2}{r^2}\!\big(x^2-3y^2\big) K_2(\kappa r)\bigg\},\nonumber\\
&\!\beta_{yx}\!=\!-\frac{b}{4\pi(1-\nu)}
\frac{x}{r^2}
\bigg\{\!(1-2\nu)+\frac{2y^2}{r^2}+\frac{4}{\kappa^2r^4}\big(x^2-3y^2\big)\nonumber\\
&\hspace{0.0cm}
-2\left((1-\nu)-\frac{y^2}{r^2}\right)\!\kappa r K_1(\kappa r)
-\frac{2}{r^2}\!\big(x^2-3y^2\big) K_2(\kappa r)\bigg\},\nonumber\\
\label{dist_yy}
&\!\beta_{yy}\!=\!-\frac{b}{4\pi(1-\nu)} 
\frac{y}{r^2}
\bigg\{\!(1-2\nu)-\frac{2x^2}{r^2}-\frac{4}{\kappa^2r^4}\big(y^2-3x^2\big)\nonumber\\
&\hspace{0.5cm}
-2\left(\frac{x^2}{r^2}-\nu\right)\kappa r K_1(\kappa r)
+\frac{2}{r^2}\big(y^2-3x^2\big) K_2(\kappa r)\bigg\},
\end{align}
and for the rotation
\begin{align}
\label{rot_z}
\omega_z\equiv-\omega=-\frac{b}{2\pi}\,\frac{x}{r^2}\Big\{1-\kappa r K_1(\kappa r)\Big\}.
\end{align}
Eq.~(\ref{rot_z}) is in agreement with the rotation vector calculated in the
linear theory of dislocations in the Cosserat continuum~\cite{Nowacki73}.
The far fields of~(\ref{dist_yy})--(\ref{rot_z}) are identical
to the classical ones given in~\cite{deWit73b}.

Now we decompose the elastic distortion~(\ref{dist_yy}) into its compatible and purely 
incompatible part by the help of Eq.~(\ref{dist2}).
Again, we may interpret the proper incompatible part as the negative plastic 
distortion. In components it reads
\begin{align}
\label{plastic-dist-edge}
\phi_{xx}=-\frac{b\kappa^2}{2\pi}\,x\varphi K_0(\kappa r),\qquad
\phi_{xy}=-\frac{b\kappa^2}{2\pi}\,y\varphi K_0(\kappa r).
\end{align}
Of course, it fulfils Eq.~(\ref{alpha-ij}).
This proper incompatible distortion is exactly the same as 
$\phi_{zx}$ and $\phi_{zy}$ of a screw dislocation (see Eq.~(\ref{plastic-dist-screw})).
The appearance of such  proper incompatible distortion is a typical 
result in elastoplasticity.
The plastic strain reads
\begin{align}
E^P_{xx}=-\phi_{xx},\qquad E^P_{xy}=-\frac{1}{2}\, \phi_{xy}.
\end{align}
The corresponding compatible part can be given in terms of 
a multi-valued displacement field
\begin{align}
\label{u_x}
&u_x=\frac{b}{2\pi}\bigg\{\big(1-\kappa r K_1(\kappa r)\big)\varphi
\nonumber\\
&\hspace{1.0cm}+\frac{1}{2(1-\nu)}\, \frac{xy}{r^2} \left(1-\frac{4}{\kappa^2r^2}+2 K_2(\kappa r)\right)\bigg\}.
\end{align}
The core radius of the edge dislocation can be estimated as $r_c\simeq 6\kappa^{-1}$.
Additionally, we find from the distortion
the incompatible part
\begin{align}
\phi_{yx}=\phi_{yy}=0,
\end{align}
and a single-valued displacement field
\begin{align}
\label{u_y}
&u_y=-\frac{b}{4\pi(1-\nu)}\bigg\{\!(1-2\nu)\big(\ln r+ K_0(\kappa r)\big)
\nonumber\\
&\hspace{1.0cm}+\frac{x^2-y^2}{2r^2}\!\left(1-\frac{4}{\kappa^2r^2}+2 K_2(\kappa r)\right)\!\bigg\}.
\end{align}
This means that $\beta_{yx}$ and $\beta_{yy}$ are proper compatible distortions.
Therefore, the field $u_y$ agrees (up to a constant term) with the
corresponding formula in gradient elasticity (see~\cite{GA97,GA99}).
The displacement field
$u_y$ differs from the classical one in the region $0\le r\le 12\kappa^{-1}$.
In the framework of elastoplasticity the displacement fields~(\ref{u_x}) and
(\ref{u_y}) have no singularity. 
Therefore, these 
displacement fields can be used to model the dislocation core. 
In this way, one can estimate the displacements, elastic strains and stresses
near the dislocation cores and compare them with HRTEM micrographs and 
atomic calculations.
The far fields of the displacements~(\ref{u_x}) and (\ref{u_y}) are identical
to the classical ones (see, e.g.,~\cite{HL}).
It is worth noting that within the Peierls-Nabarro dislocation model the 
displacement $u_y$ is identically zero.

Finally, the effective Burgers vector of the edge dislocation can be calculated as
\begin{align}
b_x(r)=\oint_\gamma\big(\beta_{xx}\d x+\beta_{xy}\d y\big)
      =b\Big\{1-\kappa r K_1(\kappa r)\Big\}.
\end{align}
Again, we find $b_x(0)=0$ and $b_x(\infty)=b$.
The effective Burgers vector $b_x(r)$ differs from the constant Burgers 
vector $b$ in the region $0\le r\le 6\kappa^{-1}$. 
Therefore, the core radius can be taken as $r_c\simeq 6\kappa^{-1}$.
For the value of $\kappa^{-1}=0.25a$ the core radius is 
$r_c=1.5a$ and for the value of $\kappa^{-1}=0.399a$ the core radius is 
$r_c=2.4a$ ($a$ is the lattice constant). 
Note that the effective Burgers vector $b_x(r)$ of an edge dislocation
has the same form as the effective Burgers vector $b_z(r)$ of a screw 
dislocation (see Eq.~(\ref{Burger-screw})).

The proper incompatible part of the elastic distortion gives rise 
to a localized dislocation density and moment stress tensor.
We find for the dislocation density of an edge dislocation
\begin{align}
\label{disl-den-edge}
\alpha_{xz}=
T_{xxy}
=\frac{b\kappa^2}{2\pi}\, K_0(\kappa r)
\end{align}
which satisfies the translational Bianchi identity~(\ref{bianchi}).
The dislocation density is short-reaching.
It is interesting to note that the dislocation density of an edge dislocation
has the same form as the dislocation density $\alpha_{zz}$ of a screw 
dislocation (see Eq.~(\ref{alpha-zz})).
In the limit as $\kappa^{-1}\rightarrow 0$, the elastoplastic
result~(\ref{disl-den-edge}) converts to the classical dislocation density
$\alpha_{xz}=b\,\delta(r)$.
The localized moment stress of bending type is given by
the help of~(\ref{moment3}) according to
\begin{align}
\label{moment-stress}
H_{xz}=\frac{\mu b}{\pi(1-\nu)}\, K_0(\kappa r),\quad
H_{zx}=-\frac{\mu b \nu}{\pi(1-\nu)}\, K_0(\kappa r).
\end{align}
This expression is very close to the moment stress of an edge dislocation
given in~\cite{HK65} (see also the remarks in~\cite{Kroener92}).

DeWit's bend-twist tensor of an edge dislocation reads
\begin{align}
&k_{zx}=\!\frac{b}{2\pi r^4}\Big\{\!\big(x^2-y^2\big)\!\big(1-\kappa r K_1(\kappa r)\big)
                -\kappa^2 x^2 r^2 K_0(\kappa r)\!\Big\},\nonumber \\
&k_{zy}=\frac{b}{2\pi r^4}\, xy \Big\{2\big(1-\kappa r K_1(\kappa r)\big)
                -\kappa^2 r^2 K_0(\kappa r)\Big\}.
\end{align}
It is in agreement with the expression calculated in the theory of Cosserat 
media (see~\cite{Kessel70,Nowacki74}).

\section{Conclusions}
\label{concl}
The field theory of elastoplasticity has been employed to consider straight
screw and edge dislocations.
The dislocation densities of straight screw and edge dislocations 
obtained in elastoplastic field theory agree with 
Eringen's two-dimensional nonlocal modulus (nonlocal kernel) used
in~\cite{Eringen83,Eringen85}.
Consequently, every component of the dislocation density tensor is Green's function of the
Helmholtz equation:
\begin{align}
\left(1-\kappa^{-2}\Delta\right)\alpha_{ij}(r)=b\,\delta(r).
\end{align}

The characteristic
internal length in elastoplasticity is $\kappa^{-1}$. This length may be
selected to be proportional to the lattice parameter $a$ 
for a single crystal, i.e.
\begin{align}
\kappa^{-1}=e_0\, a,
\end{align}
where $e_0$ is a non-dimensional constant
which can be determined by one experiment~\cite{Eringen85}.
For $e_0=0$ we recover classical elasticity.
In~\cite{Eringen83,Eringen85}
the choice $e_0=0.399$ and in~\cite{AA92} the choice 
$e_0=0.25$ are proposed. 
That length specifies the plastic region and
should be estimated by means of experimental observations and 
computer simulations of the core region.

Exact analytical solutions for the displacements, elastic and plastic strain
fields, and force stresses of dislocations have been reported which demonstrate
the elimination of any singularity at the dislocation line. 
It has been shown that the force stresses achieve their extreme values 
at a short distance away from the dislocation line.
Therefore, we are able to obtain finite strain and stress in the (linear) field
theory of elastoplasticity.
These maximum values may serve as measures of the critical stress level
of fracture.
The stress fields are in agreement with the ones obtained by Gutkin and Aifantis~\cite{GA99}.
Additionally, we have calculated the moment stresses and deWit's bend-twist tensor
for straight screw and edge dislocations.

\begin{ack}                             
The author acknowledges the support provided by the Max-Planck-Institute for Mathematics 
in the Sciences.
\end{ack}

\appendix
\section{Single-valued, discontinuous displacement and (plastic) distortion}
In the expressions given by~(\ref{u-T}), (\ref{plastic-dist-screw}), 
(\ref{plastic-dist-edge}) and (\ref{u_x})
we have considered $\varphi$ as a multi-valued field. 
On the other hand, in the defect theory~\cite{GA96,GA97,GA99,deWit73b} $\varphi$ 
is usually used as single-valued and discontinuous function. 
It is made unique by cutting the half-plane $y=0$ at $x<0$ and 
assuming $\varphi$ to jump from $\pi$ to $-\pi$ when crossing the cut.
If one uses the single-valued discontinuous form for $\varphi$,
we obtain~\cite{Lazar02c} for the screw dislocation
\begin{align}
u_z=\frac{b}{2\pi}\left\{\varphi\big(1-\kappa r K_1(\kappa r)\big)
+\frac{\pi}{2}\, {\mathrm{sign}} (y) \kappa r K_1(\kappa r)\right\},
\end{align}
with
\begin{align}
&\phi_{zx}=-\frac{b}{2\pi}\,\kappa^2 x K_0(\kappa r)
\left(\varphi-\frac{\pi}{2}\, {\mathrm{sign}}(y)\right),\nonumber\\
&\phi_{zy}=-\frac{b}{2\pi}
\Big\{\kappa^2 y K_0(\kappa r)\left(\varphi-\frac{\pi}{2}\, {\mathrm{sign}}(y)\right)
\nonumber\\
&\hspace{1.0cm}+\pi\delta(y)\Big(1-{\mathrm{sign}}(x)
\big[1-\kappa r K_1(\kappa r)\big]\Big)\Big\},
\end{align}
and for the edge dislocation
\begin{align}
&u_x=\frac{b}{2\pi}\bigg\{\varphi\big(1-\kappa r K_1(\kappa r)\big)
+\frac{\pi}{2}\, {\mathrm{sign}} (y)\,\kappa r K_1(\kappa r)\nonumber\\
&\hspace{1.0cm}
+\frac{1}{2(1-\nu)}\, \frac{xy}{r^2}\left(1-\frac{4}{\kappa^2r^2}+2 K_2(\kappa r)\right)\bigg\},\nonumber\\
&u_y=-\frac{b}{4\pi(1-\nu)}\bigg\{(1-2\nu)\big(\ln r+ K_0(\kappa r)\big)
\nonumber\\
&\hspace{1cm}+\frac{x^2-y^2}{2r^2}\left(1-\frac{4}{\kappa^2r^2}+2 K_2(\kappa r)\right)\bigg\},
\end{align}
with
\begin{align}
&\phi_{xx}=-\frac{b}{2\pi}\,\kappa^2 x K_0(\kappa r)
\left(\varphi-\frac{\pi}{2}\, {\mathrm{sign}}(y)\right),\\
&\phi_{xy}=-\frac{b}{2\pi}
\label{plastic-dist2-y}
\Big\{\kappa^2 y K_0(\kappa r)\left(\varphi-\frac{\pi}{2}\, {\mathrm{sign}}(y)\right)
\nonumber\\
&\hspace{1cm}+\pi\delta(y) \Big(1-{\mathrm{sign}}(x)\big[1-\kappa r K_1(\kappa r)\big]\Big)\Big\},\nonumber\\
&\phi_{yx}=\phi_{yy}=0.\nonumber
\end{align}
For a more detailed discussion see~\cite{Lazar02c}.


\end{document}